\documentclass[aps,pre,twocolumn,showpacs,superscriptaddress,groupedaddress]{revtex4-1}  % for review and submission
\usepackage{graphicx}  % needed for figures
\usepackage{dcolumn}   % needed for some tables
\usepackage{bm}        % for math
\usepackage{amssymb}   % for math
\usepackage{amsmath}   % for math

\begin{document}

\title{Understanding plastic deformation in thermal glasses from single-soft-spot dynamics}
\author{S. S. Schoenholz}
\email{schsam@sas.upenn.edu}
\affiliation{Department of Physics, University of Pennsylvania, Philadelphia, Pennsylvania 19104, USA}
\author{A. J. Liu}
\affiliation{Department of Physics, University of Pennsylvania, Philadelphia, Pennsylvania 19104, USA}
\author{R. A. Riggleman}
\affiliation{Department of Chemical and Biomolecular Engineering, University of Pennsylvania, Philadelphia, PA 19104, USA }
\author{J. Rottler}
\affiliation{Department of Physics and Astronomy, University of British Columbia, Vancouver, BC V6T1Z4, Canada}
\date{\today}

\begin{abstract}
By considering the low-frequency vibrational modes of amorphous solids, Manning and Liu [Phys. Rev. Lett. \textbf{107}, 108302 (2011)] showed that a population of ``soft spots'' can be identified that are intimately related to plasticity at zero temperature under quasistatic shear. In this work we track individual soft spots with time in a two-dimensional sheared thermal Lennard Jones glass at temperatures ranging from deep in the glassy regime to above the glass transition temperature. We show that the lifetimes of individual soft spots are correlated with the timescale for structural relaxation. We additionally calculate the number of rearrangements required to destroy  soft spots, and show that most soft spots can survive many rearrangements. Finally, we show that soft spots are robust predictors of rearrangements at temperatures well into the super-cooled regime. Altogether, these results pave the way for mesoscopic theories of plasticity of amorphous solids based on dynamical behavior of individual soft spots.
\end{abstract}

\maketitle

\section{Introduction} 

Solids flow under shear via localized rearrangements. In crystals it is known that this flow is achieved via the propagation of topological defects~\cite{dislocations}. In disordered systems, flow is also achieved via rearrangements that occur at localized regions, but it has proven difficult to locate the regions in advance of the rearrangements.  One way of identifying them is via their ability to scatter sound waves.  Regions that are particularly effective in scattering sound appear as regions of high polarization in low-frequency quasi localized vibrational modes.  The high-polarization regions have been shown to be vulnerable to rearrangement under applied stress or temperature~\cite{widmer08,cooper09,candelier10,hocky13,manning11,chen11,chen13,rottler14}.  
Manning and Liu~\cite{manning11} therefore used low-frequency quasi localized modes to construct a population of localized regions, or ``soft spots,'' which they showed were highly correlated with rearrangements induced by quasi static shear at zero temperature.
%More recently, it has been shown that not only do soft spots capture flow defects in systems ranging from nearly perfect crystals with isolated dislocations to fully disordered packings~\cite{chen11,chen13,joerg13} but that the directions of particle displacements in the relevant quasi localized modes are highly correlated with the directions of particle displacements in rearrangements~\cite{joerg13}.

One promising theoretical approach to plasticity in glasses has been to construct a mesoscopic phenomenological theory based on a population of localized structural flow defects, or regions of enhanced fluidity, that are prone to rearrangement.  This is the approach adopted by shear transformation zone theory~\cite{falk98,falk11} and by mesoscopic kinetic elastoplasticity models~\cite{bocquet09,mansard11}.  Soft spots are obvious candidates for the flow defects that lie at the heart of these models.  In order for soft spots to serve as a useful basis for a mesoscopic theory of plastic flow, however, two minimal conditions must be met.  

First, rearrangements must preferentially occur at soft spots, not only at zero temperature under quasi static shear, but at temperatures extending at least to the glass transition temperature, and realistic shear rates. Here we show that soft spots do indeed correlate with rearrangements at temperatures ranging from well below the glass transition to above the transition, over a range of shear rates, in two-dimensional model glasses.

Second, soft spots must survive long enough for their dynamics to capture the slow relaxation time of a sheared glassy system.  In this paper, we track individual soft spots with time. We show that the average lifetime of soft spots correlates with the relaxation time of a glass. Surprisingly, most soft spots can withstand many rearrangements before being destroyed. This longevity leads to a distribution of soft spot lifetimes that follows a power-law up to the structural relaxation time. 

Together, these two main conclusions provide strong support for a mesoscopic approach to plasticity in glasses that is based on dynamics of the soft spot population.

In section II we describe how we study soft spots and their correlation with rearrangements in sheared thermal glasses.  Section III shows that soft spots obtained from inherent structures correlate well with rearrangements that follow in a short interval of time later.   The degree of correlation decreases with temperature, but soft spots remain a valid description of plastic activity in amorphous solids at temperatures ranging from those deep in the glassy phase up through the glass transition.  Section IV describes how the soft spot population decorrelates with time on a time scale comparable to the relaxation time, obtained from the decay of the intermediate scattering function.   In Section V, we turn to the dynamics of individual soft spots and show that the decorrelation of the soft spot population can be understood in terms of the single soft-spot dynamics.   These results demonstrate the deep and robust connection between soft spots and plasticity in amorphous matter.

\section{Methods} 

To study the effects of temperature and strain rate on the validity of the soft spot picture, we consider a 10,000-particle, two-dimensional, 65:35 binary Lennard-Jones mixture. We use a model with the parameters $\sigma_{AA} = 1.0$, $\sigma_{AB} = 0.88$, $\sigma_{BB} = 0.8$, $\epsilon_{AA} = 1.0$, $\epsilon_{AB} = 1.5$, and $\epsilon_{BB} = 0.5$. The Lennard-Jones potential is cut off at $2.5\sigma_{AA}$ and smoothed so that both first and second derivatives go continuously to zero at the cutoff. The natural units for the simulation are $\sigma_{AA}$ for distances, $\epsilon_{AA}$ for energies, and $\tau = \sqrt{m\sigma_{AA}^2/\epsilon_{AA}}$ for times. We perform molecular dynamics simulations of this system using LAMMPS with a timestep of $5\times 10^{-3}\tau$ at density $\rho=1.2$. A Nos\'e-Hoover thermostat with a time constant of $1 \tau$ is used to keep the system at a fixed temperature. We consider temperatures $T = 0.1, 0.2, 0.3,$ and $0.4$ as well as strain rates $\dot\gamma = 10^{-5}, 10^{-4},$ and $10^{-3}$. In all cases data was collected after allowing the system to reach steady state by shearing up to 20\% strain. This system has been characterized and shown to be a good glass former by Br\"uning \textit{et al.}~\cite{bruning09}. Notably, it was shown~\footnote{The glass transition temperature was determined by constructing a fictive temperature~\cite{tool46}, $T_f(T)$, that is equal to the real temperature of the system in the liquid phase, but approaches a constant as the system falls out of equilibrium. Thus $T_g$ is defined to be asymptotic value, $T_f(0)$.} that the glass transition temperature for this model is $T_G = 0.33$.  Therefore, at the highest temperature we are studying a system well into the supercooled regime.

\begin{figure}[!ht]
\includegraphics[width=0.48\textwidth]{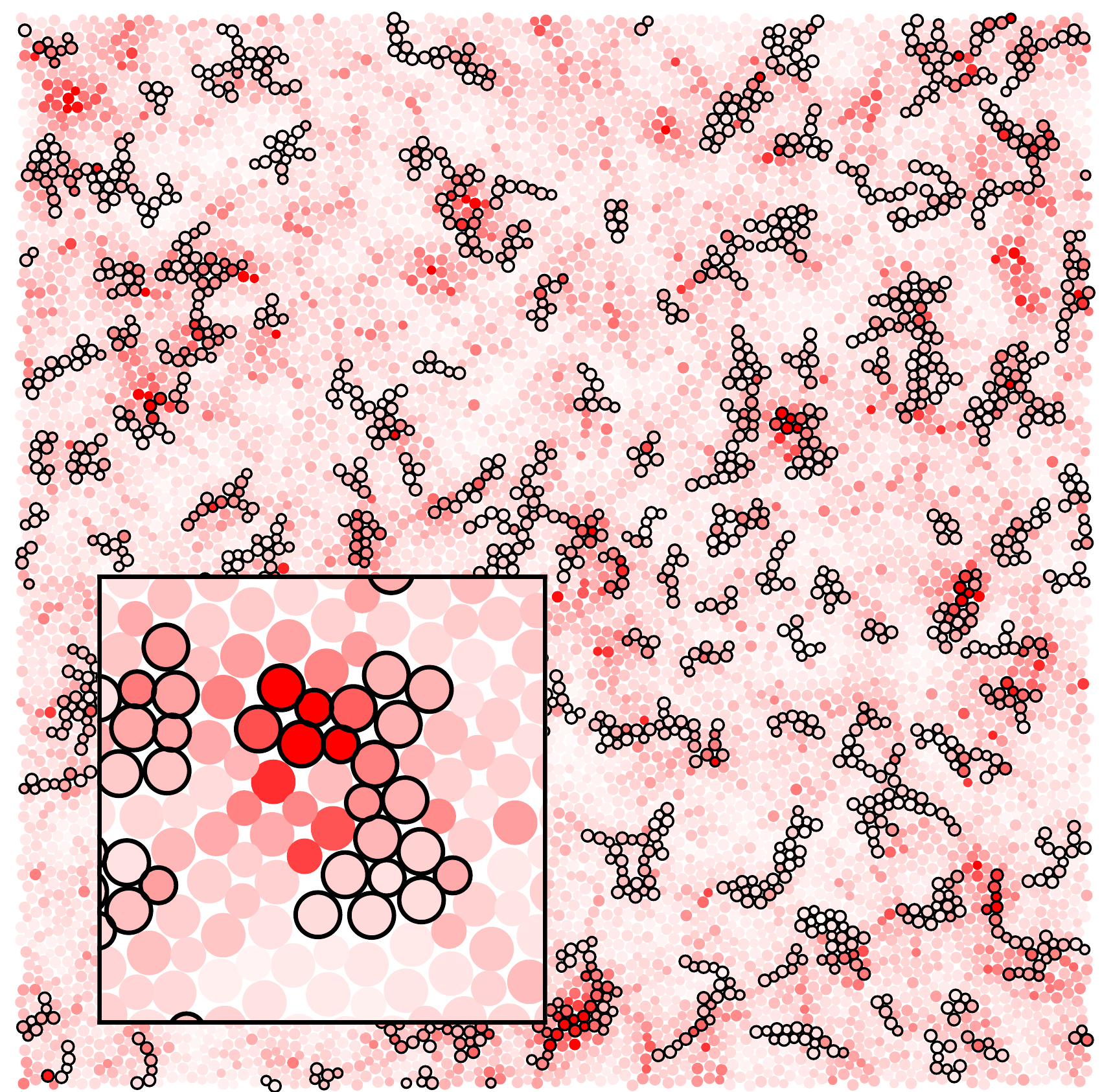}
\caption{An example configuration of the system at $T=0.1$ and $\dot\gamma=10^{-4}$. Particles are colored according to their $D^2_{\text{min}}$ value (see text). Particles outlined in black are members of the soft-spots for this configuration. The soft-spots have been generated using $N_m = 430$ and $N_p = 20$. Inset: a single soft spot coinciding with a rearrangement.}
\label{fig:system}
\centering
\end{figure}

To construct the soft spots we begin with a harmonic description of the inherent structure of the glass and follow the procedure of Manning and Liu~\cite{manning11}. Therefore, every $2\tau$ we quench the system to its inherent configuration using a combination of the conjugate gradient and FIRE algorithms~\cite{bitzek06}. We then compute the 500 lowest frequency modes by diagonalizing the dynamical matrix using ARPACK \footnote{ARPACK can be found at www.caam.rice.edu/software/ARPACK.}. The boson peak for this system occurs, on average, at 270 modes; therefore, this set of modes captures the low-frequency harmonic behavior of the system. From this collection we select the $N_m$ most localized modes ranked by their participation ratios\cite{xu10}. From these $N_m$ modes we further select the $N_p$ particles with the largest polarization vectors.  The parameters $N_m$ and $N_p$ are not free, but are rather chosen to maximize the correlation of the soft-spot population with rearrangements. The details of the selection will be discussed below.  Finally, we remove clusters of fewer than four particles since at least 4 particles are required for a T1 rearrangement. An example of the soft spot population is shown in fig.~\ref{fig:system}.  We emphasize that, as found by Manning and Liu~\cite{manning11}, the qualitative results presented in this paper are remarkably insensitive to the details of the protocol used to select the modes, the choices of $N_m$ and $N_p$, as well as the choice to remove small clusters.  Changes of this sort affect the magnitude of the correlations that we present, but will not affect the existence, duration, or even functional forms of these correlations.

Given a set of particles comprising our soft spot population, we can then construct an $N$ dimensional projection operator $\bm S(t)$ so that $S_i(t) = 1$ if particle $i$ is in a soft spot and $S_i(t) = 0$ otherwise. Additionally, we define the overall fraction of space covered by a soft spot to be $\rho_{SS} =  \langle S_i(t)\rangle$ where the average is taken over particles and times. 
We measure the plastic rearrangements of the system using the quantity $D^2_{\text{min}}$ as introduced by Falk and Langer \cite{falk98}. $D^2_{\text{min}}$ is defined to be the amount of locally non-affine displacement that particles undergo in a time interval $\Delta t$. A value of $D^2_{\text{min}}$ can be associated with each particle,
\begin{align}
D_i(t,\Delta t) = \sum_j\big[&\bm r_j(t+\Delta t) - \bm r_i(t + \Delta t)\nonumber \\
&- \bm\Lambda_{i}(\bm r_j(t) - \bm r_i(t))\big]^2
\end{align}
where the sum is taken over particles in a local neighborhood to particle $i$ and $\Lambda_{i}$ is the affine transformation that minimizes $D_i$. In our study we use neighborhoods of $2.5\sigma_{AA}$ to be the same size as the Lennard-Jones cutoff. Additionally we use $\Delta t = 2\tau$ to be the same as the scale on which we generate soft spot configurations. Our analysis has been repeated for various values of $\Delta t$ and has proven to be insensitive to its value, as long as it is on the same order as the duration of a plastic event. Fig.~\ref{fig:system} shows - in addition to the soft spot population - a map of the local $D^2_{\text{min}}$ amplitudes.  Darker regions indicating higher values of $D^2_{\text{min}}$ tend to lie on top of soft spots, showing that indeed rearrangements occur preferentially at soft spots.

\section{Equal-time correlations} 

\begin{figure}
\hspace{-0.75pc}\includegraphics[width=0.5\textwidth]{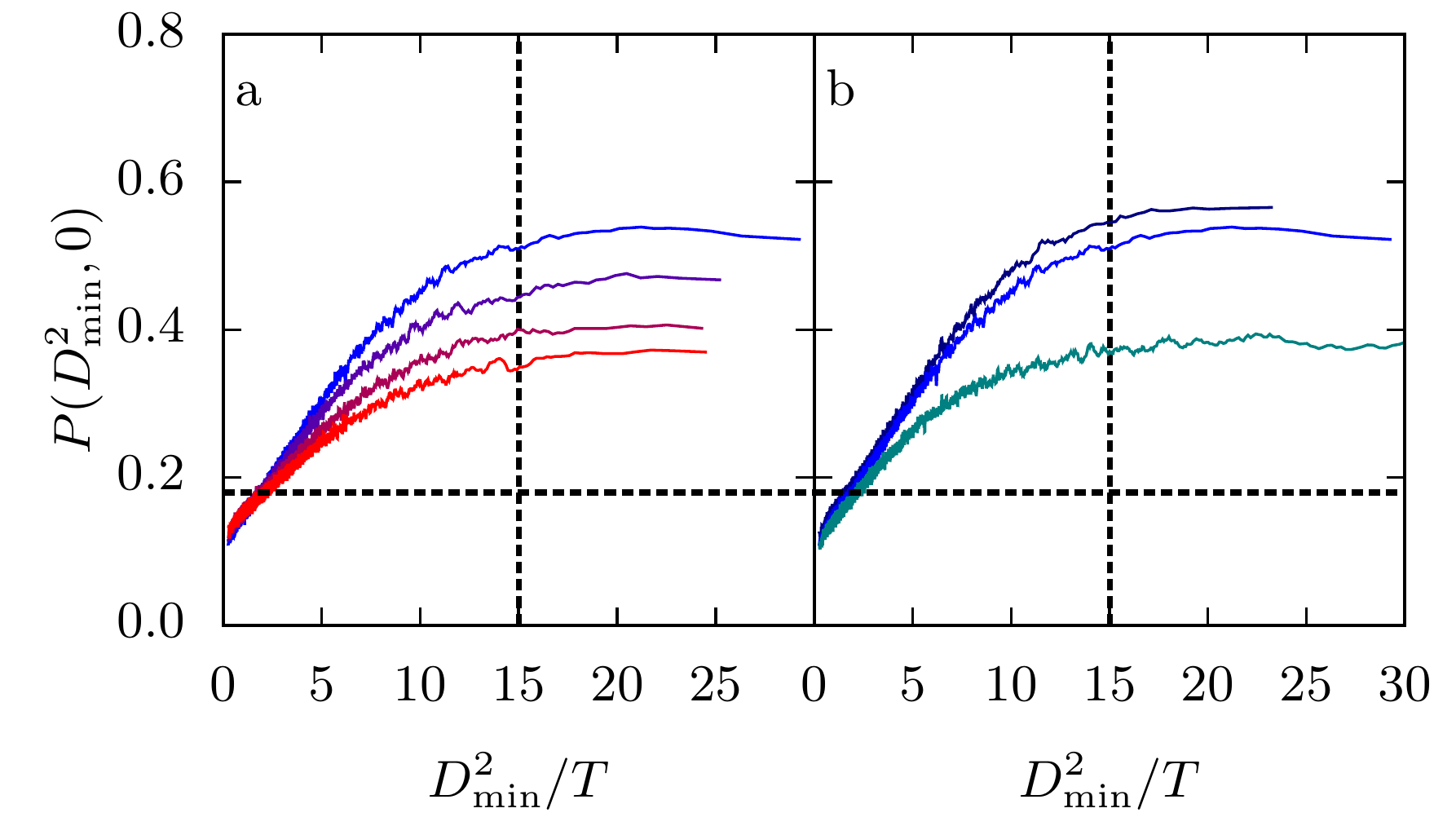}
\caption{The probability of a particle residing in a soft-spot as a function of its $D^2_{\text{min}}$ value. (a) shows a comparison of the temperatures studied from $T=0.1$ in blue to $T=0.4$ in red at a strain rate of $\dot\gamma=10^{-4}$ and (b) shows a comparison of strain rates studied from $\dot\gamma = 10^{-5}$ in dark blue to $\dot\gamma = 10^{-3}$ in green at a temperature of $T = 0.1$. In all cases we see that the probability increases with $D^2_{\text{min}}$ until some threshold $D^2_{\text{th}}\simeq 15T$ (vertical dashed line) at which point the probability reaches some plateau value $P^*$. The soft spot density, $\rho_{SS}$, is marked by a horizontal dashed line. }
\label{fig:overlap}
\centering
\end{figure}

In order to quantify the degree of correlation between soft spots and plasticity, we consider the probability, $P(D^2_{\text{min}})$, that a particle with a given $D^2_{\text{min}}$ value in the interval $[t,t+\Delta t]$ resides in a soft spot constructed from the inherent structure at a time $t$.  Thus, we study the correlations of soft spots at time $t$ with rearrangements characterized during a short time interval $\Delta t$ following $t$.  This may be expressed as 
\begin{equation}
P(D^2_{\text{min}}) = \frac{\langle\delta(D_i(t)-D^2_{\text{min}})S_i(t)\rangle}{\langle\delta(D_i(t)-D^2_{\text{min}})\rangle}.
\end{equation}
If the soft spot map is uncorrelated with the $D^2_{\text{min}}$ map, then this quantity simply reduces to the soft spot density, $\rho_{SS}$, independent of $D^2_{\text{min}}$. This equal-time probability is shown in fig.~\ref{fig:overlap} for four temperatures and three strain rates. In all cases, we see that $P(D^2_{\text{min}})$, rises to a plateau value as $D^2_{\text{min}}$ increases. Therefore it is clear that particles with higher $D^2_{\text{min}}$ are more likely to reside in soft spots. Conversely, particles with very small values of $D^2_{\text{min}}$ appear to be anti correlated with soft spots, since $P(D^2_{\text{min}})$ is smaller than $\rho_{SS}$.  

The plateau value of $P(D^2_{\text{min}})$, $P^*(T,\dot\gamma)$, decreases with increasing temperature and strain rate. Thus, the descriptive power of the soft spot picture is reduced by increasing temperature or strain rate, as expected.  In order to compare results for different temperatures we divide $D^2_{\text{min}}$ by $T$,  since the $D^2_{\text{min}}$ of particles not undergoing rearrangements is due to thermal fluctuations in which case $D^2_{\text{min}} \sim \langle v^2\rangle\sim T$ by the equipartition theorem. The probability appears to reach its plateau value for $D^2_{\text{min}} \gtrsim 15T$ independent of strain rate.  We therefore define the plateau probability, $P^*(T,\dot\gamma)$, to be a good measure of the equal time correlation between the $D^2_{\text{min}}$ map and the soft spot map at a given temperature and strain rate. 

%Finally, at low $D^2_{\text{min}}$ the probability appears to collapse with $D^2_{\text{min}}\to D^2_{\text{min}}/T$. 
%As expected as $\delta t\to\infty$ the soft spot map becomes uncorrelated with the $D^2_{\text{min}}$ map and we find that $P^*(\delta t,T,\dot\gamma)\to\rho_{SS}$.

\begin{figure}
\includegraphics[width=0.45\textwidth]{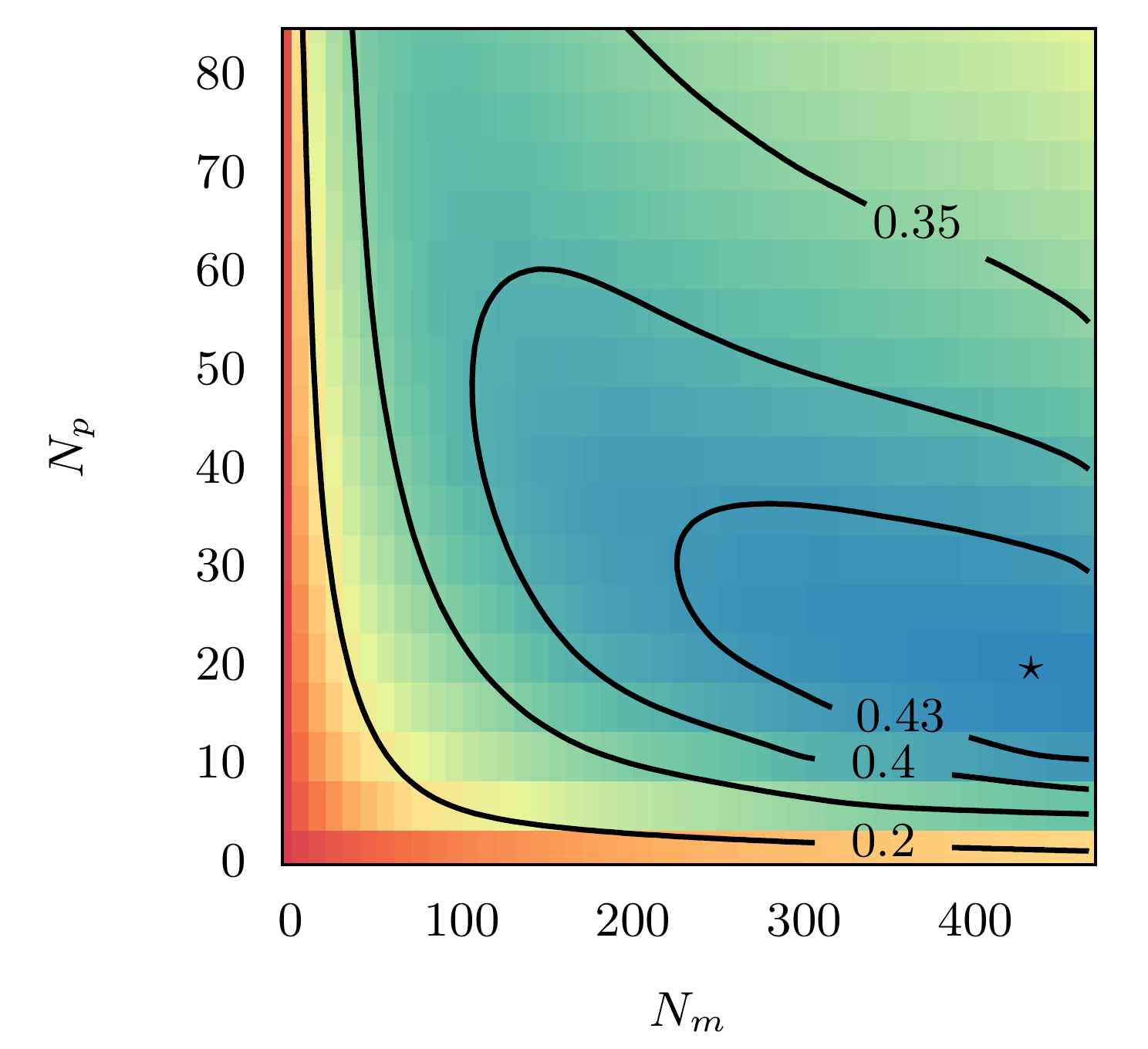}
\caption{The difference in probability, $\Delta P^* = P^*(T,\dot\gamma) - \rho_{SS }$, as a function of $N_m$ and $N_p$ for a temperature of $T = 0.1$ and strain rate $\dot\gamma = 10^{-4}$. We see a broad plateau over which $\Delta P^*$ is largely independent of $N_m $ and $N_p$ with a weak maximum occurring at $N_m^\star = 430$ and $N_p^\star = 20$ (marked by a star.) The behavior of $\Delta P^*$ is largely independent of temperature and strain rate.}
\label{fig:correlationcontour}
\centering
\end{figure}

We now discuss the choice of $N_m$ and $N_p$. Following Manning and Liu~\cite{manning11}, we select $N_m$ and $N_p$ to maximize the correlation between the soft spots and the $D^2_{\text{min}}$ map. To this end, we consider the difference, $\Delta P^* = P^*(T,\dot\gamma) - \rho_{SS}$, as a function of $N_m$ and $N_p$.  Recall that $\rho_{SS}$ represents the value of $P^*(T,\dot\gamma)$ if the soft-spot map is uncorrelated with the $D^2_{\text{min}}$ map. Thus, adding particles to the soft spot map that are correlated with the $D^2_{\text{min}}$ map will increase $P^*(T,\dot\gamma)$ more than $\rho_{SS}$; conversely, adding particles to the soft spot map that are anti-correlated with the $D^2_{\text{min}}$ map will increase $P^*(T,\dot\gamma)$ less than $\rho_{SS}$. Therefore, a maximum in $\Delta P^*(T,\dot\gamma)$ at some $N_m^\star$ and $N_p^\star$  represents the selection of parameters that yields the maximal correlation. As shown in fig.~\ref{fig:correlationcontour}, we find a broad plateau as a function of $N_m$ and $N_p$ with a maximum at $N_m^\star = 430$ and $N_p^\star = 20$.  We do not see a strong dependence of $\Delta P^*$ on either temperature or strain rate, so we use these values at all temperatures and strain rates studied.

\begin{figure}[!h]
\includegraphics[width=0.45\textwidth]{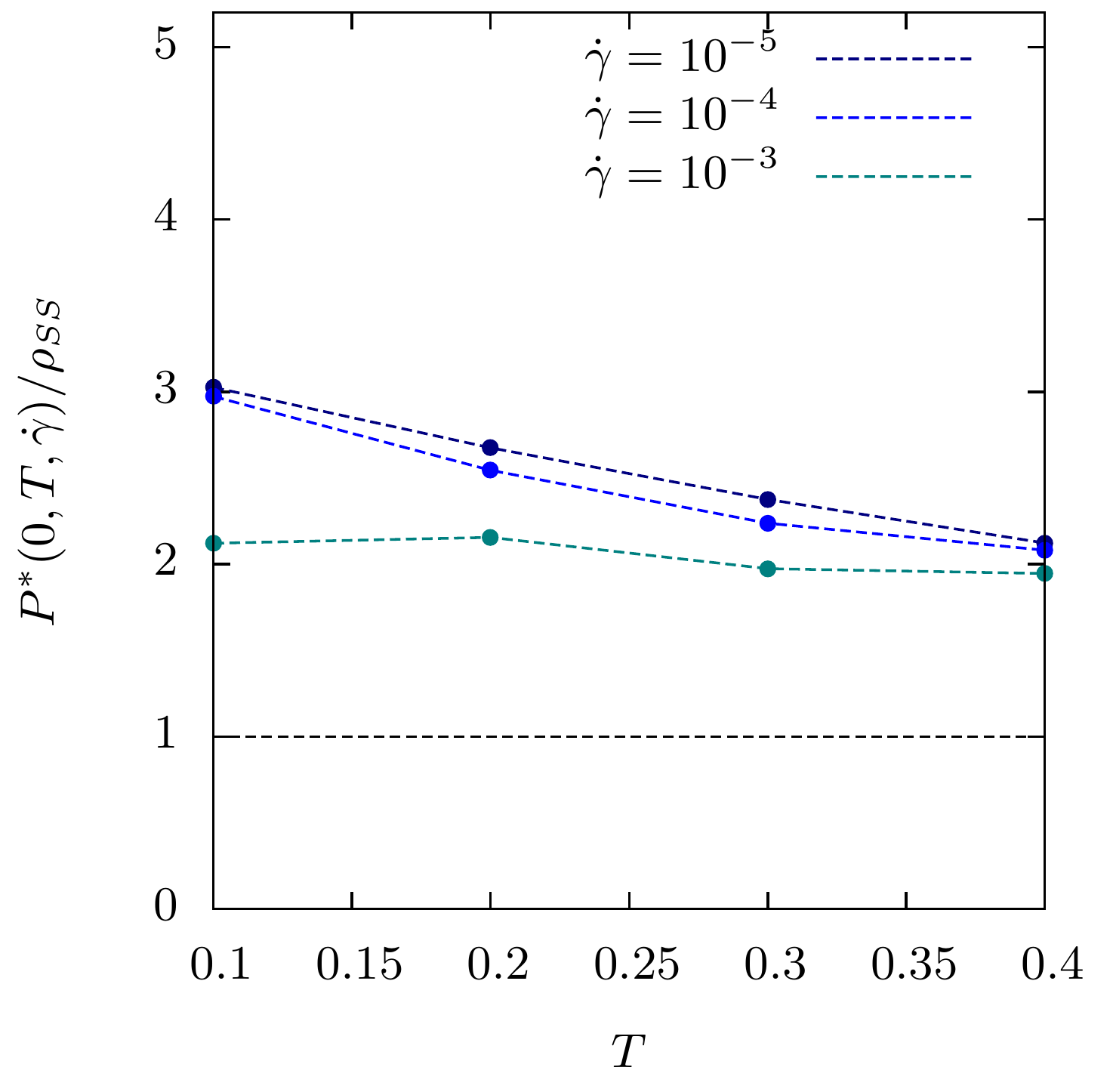}
\caption{The plateau probability, $P^*$, for a particle with high $D^2_{\text{min}}$ to reside in a soft spot, normalized by the soft spot density, $P_{SS}$. This represents how much more likely rearrangements are to be found at soft spots than if the soft spot map were completely uncorrelated with rearrangements. A value of 1 (dashed line) represents the uncorrelated value. The ratio is 0 when the soft spot map is anti-correlated and so describes no plastic activity. The value of 5.2 represents maximum possible value of $P^*$, $1/\rho_{SS}$, which occurs if all of the plastic activity resides in soft spots. Data is shown for strain rates $\dot\gamma = 10^{-5}$ in dark blue to $\dot\gamma = 10^{-3}$ in green.}
\label{fig:crosscorrelation}
\centering
\end{figure}

Having identified a population of soft spots, we now discuss the degree to which soft spots correlate with plastic activity over the temperatures and strain rates studied. To understand this correlation we consider the ratio $P^*(T,\dot\gamma)/\rho_{SS}$. This ratio measures how much more likely rearrangements are to occur on soft spots than on a randomly distributed set of particles at the same density. Since there are always more soft spots than rearranging particles we have the bounds, $0 \leq P^*(T,\dot\gamma)/\rho_{SS} \leq 1/\rho_{SS}$. The upper bound occurs if every rearranging particle resides on a soft spot. If the soft spot map and the $D^2_{\text{min}}$ map were uncorrelated we would expect $P^*(T,\dot\gamma)/\rho_{SS} = 1$.  We plot this ratio in fig.~\ref{fig:crosscorrelation} as one of the central results of this paper. We see the highest correlation at the lowest temperature and rate, where plastic events are more than three times as likely to occur at soft spots than if the soft spots were randomly distributed. As temperature and strain rate are increased, these correlations decrease slightly. Even at the highest temperature and strain rate, well above the glass transition temperature for the system, we continue to see that rearrangements are twice as likely to occur at soft spots than they would be in the rearrangements were randomly distributed in the system. We conclude that soft spots are robust in describing plastic activity in glassy materials under shear, not only at low temperatures but also well into the supercooled regime.

\section{Time-dependent correlations}

We now characterize the time dependence of various correlations. We will show that two soft spot maps, constructed a time $\delta t$ apart, remain correlated up to the longest timescale for relaxation in the system, the $\alpha$-relaxation time. We will further demonstrate that the decorrelation of these two soft spot configurations is approximately logarithmic in time. Finally we observe that this behavior is mirrored in the autocorrelation function of the $D^2_{\text{min}}$ field with itself as well as in the crosscorrelation function between the $D^2_{\text{min}}$ field and the soft spot map. We conclude that soft spots represent long-lived structural features of glassy systems that are intimately related to flow and failure of these materials.

The $\alpha$-relaxation time, $\tau_\alpha$, is a measure of the amount of time needed for every particle in the system to experience a rearrangement. A common method for defining $\tau_\alpha$ is via the decay of the self part of the intermediate scattering function,
 $F_s(\bm q,\delta t) = \langle\exp[i\bm q\cdot(\bm r_i(t+\delta t) - \bm r_i(t))]\rangle$. The $\alpha$-relaxation time can be effectively defined to be the time at which $F_s(\bm q_{\text{max}},\delta t) \sim e^{-1}$ where $\bm q_{\text{max}}\simeq 2\pi/\sigma_{AA}\bm{\hat y}$ is the wave-vector at the first maximum of the static structure factor. Here we take $\bm q_{\text{max}}$ orthogonal to the axis of imposed shear to avoid artifacts from the affine component of displacement. 
 
\begin{figure}[!ht]
%\hspace{2.5pc}\input{D2Min_Correlations.tex}\vspace{2.0pc}
\hspace{-0.55pc}\includegraphics[width=0.48\textwidth]{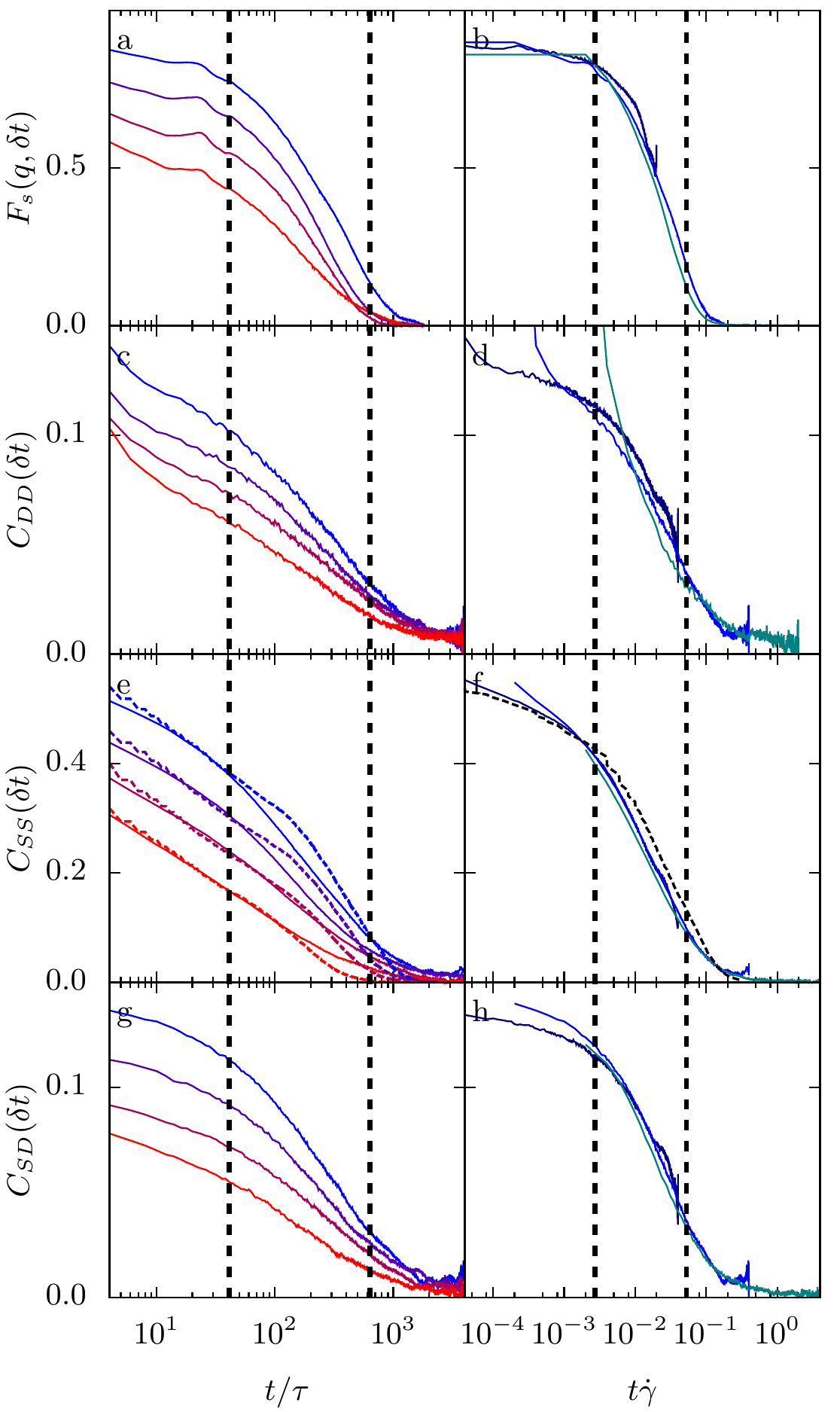}\vspace{-1.0pc}
\caption{Correlation functions of the $D^2_{\text{min}}$ field and the soft-spot population. On the left are comparisons of the temperatures considered from blue for $T=0.1$ to red for $T=0.4$ at a strain rate of $\dot\gamma=10^{-4}$.  On the right are comparisons of the strains considered from dark blue for $\dot\gamma = 10^{-3}$ to green for $\dot\gamma=10^{-5}$ scaled by the strain, $t/\tau\to\dot\gamma t$ at a temperature $T = 0.1$. Figures (a)-(b) show the self-intermediate scattering function, $F(q_{\text{max}},t)$, evaluated at $\bm q = 2\pi/\sigma_{AA}\bm{\hat y}$. Figures (c)-(d) show the autocorrelation function for $D^2_{\text{min}}$. Figures (e)-(f) show the autocorrelation function for the soft spot population. In (e) at each temperature comparisons are made with the cumulative probability density function for individual soft spot lifetimes, introduced in section V, overlaid in dashed lines. In (f) a single comparison is made to $P(\tau_L\geq\delta t)$, shown using a dashed black line, for lifetimes aggregated from lifetimes collected at all three different strain rates. Figures (g)-(h) show the cross correlation between $D^2_{\text{min}}$ and the soft spot population. On each side there are two vertical dashed lines to serve as guides to the eye. The earlier line occurs at a time, $\tau^*$, when the self-intermediate scattering function first drops below the plateau. The later line occurs at the $\alpha$-relaxation time, $\tau_\alpha$, defined so that $F(q_{\text{max}},\tau_\alpha) \sim e^{-1}$. }
\label{fig:correlation}
\centering
\end{figure}

 A plot of the self-intermediate scattering function for a range of temperatures, at a strain rate of $\dot\gamma = 10^{-4}$, and shear rates, at a temperature of $0.1$, can be seen in fig.~\ref{fig:correlation} (a) and (b) respectively. We see that at short times, $F_s(q_{\text{max}},\delta t)$ falls to some plateau - whose value decreases with temperature - before dropping precipitously to zero. The time at which $F_s(q_{\text{max}},\delta t)$ first deviates from the plateau, which we shall denote $\tau^*$, appears to feature only weak temperature dependence. For this shear rate and the temperatures shown, $\tau^* \approx 50 \tau$.  We further notice that the self-intermediate scattering function collapses with $\delta t/\tau \to \dot\gamma\delta t$ which indicates that $\tau_\alpha\sim\dot\gamma$ for the strain rates considered. We therefore conclude that at this low temperature, we are in a regime where strain dominates the plastic flow of the system. 

We now quantify the time-dependent correlation functions. As is customary we define the correlation function for two fields $X_i(t)$ and $Y_i(t)$ by,
\begin{equation}
C_{XY}(\delta t)  = \langle\tilde X_i(t+\delta t)\tilde Y(t)\rangle.
\end{equation} 
Here $\tilde X_i(t)$ and $\tilde Y_i(t)$ are fields constructed from $X_i(t)$ and $Y_i(t)$ normalized to have zero mean and unit variance. The autocorrelation functions, $C_{SS}(\delta t)$ and $C_{DD}(\delta t)$, as well as the cross correlation function $C_{SD}(\delta t)$ can be seen in fig.~\ref{fig:correlation} (c)-(h). Examining the figures, we notice first and foremost that all of the correlation functions decay to zero with $F_s(q_{\text{max}},\delta t)$. We therefore conclude that the $D^2_{\text{min}}$ and soft spot fields remain correlated up to the longest time scale for relaxation in the system, the $\alpha$-relaxation time, at all temperatures and strain rates considered. We note that the $C_{SD}$ function has a modest equal time correlation $C_{SD}(0)\sim 0.14$ due to the fact that there are far more soft spots than rearranging particles. This is a problem that we circumvented in the preceding section by using the quantity $P^*$ to quantify equal-time correlations. 

Examining fig.~\ref{fig:correlation} further, we notice that the correlation functions are all qualitatively similar. In each case the functions experience a drop at a timescale shorter than our time resolution of $2\tau$, followed by a slow decay up until $\tau^*$, at which point they decrease quickly, dropping to zero at approximately $\tau_{\alpha}$. The initial drop appears to bring the correlation function to a value that depends on temperature but not strain rate. The correlation functions appear to collapse as $\delta t \to \dot\gamma\delta t$ which reinforces our conclusion that we are in a regime where plastic flow is controlled by strain. The exception to this collapse is in the short-time behavior of the $D^2_{\text{min}}$ autocorrelation function. Here we see that the functions collapse at times greater than some strain rate independent cutoff time; however, at shorter times, they increase quickly from this master curve. We attribute this fast increase at short times to an exponential decorrelation of the $D^2_{\text{min}}$ map due to the finite duration of plastic events. The soft spots feature no such effect since they are constructed from the inherent structure of the system.

We conclude that two soft spot maps, constructed a time $\delta t$ apart, remain correlated until almost every particle in the system has experienced a rearrangement. This remarkable stability of the soft spot configuration appears to be robust to increasing temperatures and strain rates. Furthermore, the extremely slow decorrelation is mirrored in the dynamics of the system, suggesting that plastic activity in glassy systems is intimately tied to structural soft spots. Finally, note that the correlation functions are qualitatively similar, not only to one another, but also to the self-intermediate scattering function itself. Therefore an improved understanding of the decorrelation of the soft spot configuration might shed some light on structural relaxations in glassy systems. 

\section{Single-soft-spot dynamics}

We now decompose the behavior of the soft spot field as a whole in terms of the dynamics of individual soft spots. In particular we will show that the form of the soft spot autocorrelation function seen in fig.~\ref{fig:correlation} (e)-(f) can be explained by understanding the lifetime of individual soft spots in the configuration. To do this we will first construct upper bounds on single-soft-spot lifetimes. We will then introduce distributions of single-soft-spot lifetimes over a range of temperatures and strain rates. Finally, we will argue that the $C_{SS}(\delta t)$ function arises naturally from the distribution of these lifetimes. We will conclude by considering the number of rearrangements necessary to destroy an individual soft spot. By introducing a simple model with no correlations, we show that the distribution of single-soft-spot lifetimes can be estimated directly from this latter quantity.

We construct an upper bound on single-soft-spot lifetimes by considering the autocorrelation function of individual soft spots. To do this we first consider the fields $\bm S_{\alpha}(t)$, for an individual soft spot labelled $\alpha$, defined so that $S_{\alpha,i}(t) = 1$ if particle $i$ is in soft spot $\alpha$ and $S_{\alpha,i}(t) = 0$ otherwise. To determine how a soft spot, $\alpha$, constructed at a time $t$ evolves at a time $t+\delta t$, we define an autocorrelation function,
\begin{equation}\label{eq:single}
C_{S_\alpha S_\alpha}(\delta t) = \max_\beta\langle\tilde S_\beta(t+\delta t)\tilde S_\alpha(t)\rangle.
\end{equation}
Eq.~\eqref{eq:single} associates a soft spot at time $t$ with the ``best'' soft spot at a time $t+\delta t$. We then average $C_{S_\alpha S_\alpha}(\delta t)$ over a moving window of width $\Delta t = 16\tau$ to remove some of the fluctuations. The lifetime of a soft spot, denoted $\tau_L$, is then defined as the first time at which the averaged autocorrelation function dips below its asymptotic, uncorrelated, value,
\begin{equation}
\varepsilon = \lim_{\delta t\to\infty}\langle C_{S_\alpha S_\alpha}(\delta t)\rangle_S\approx0.2
\end{equation}
where $\langle\cdot\rangle_S$ denotes an average over soft spots.  We have found the results to be qualitatively insensitive to the choice of either $\Delta t$ or $\varepsilon$.  Note that since we have used the maximum function in eq.~\ref{eq:single}, $\tau_L$ represents an upper bound on the actual soft spot lifetime. 

\begin{figure}[!ht]
\includegraphics[width=0.5\textwidth]{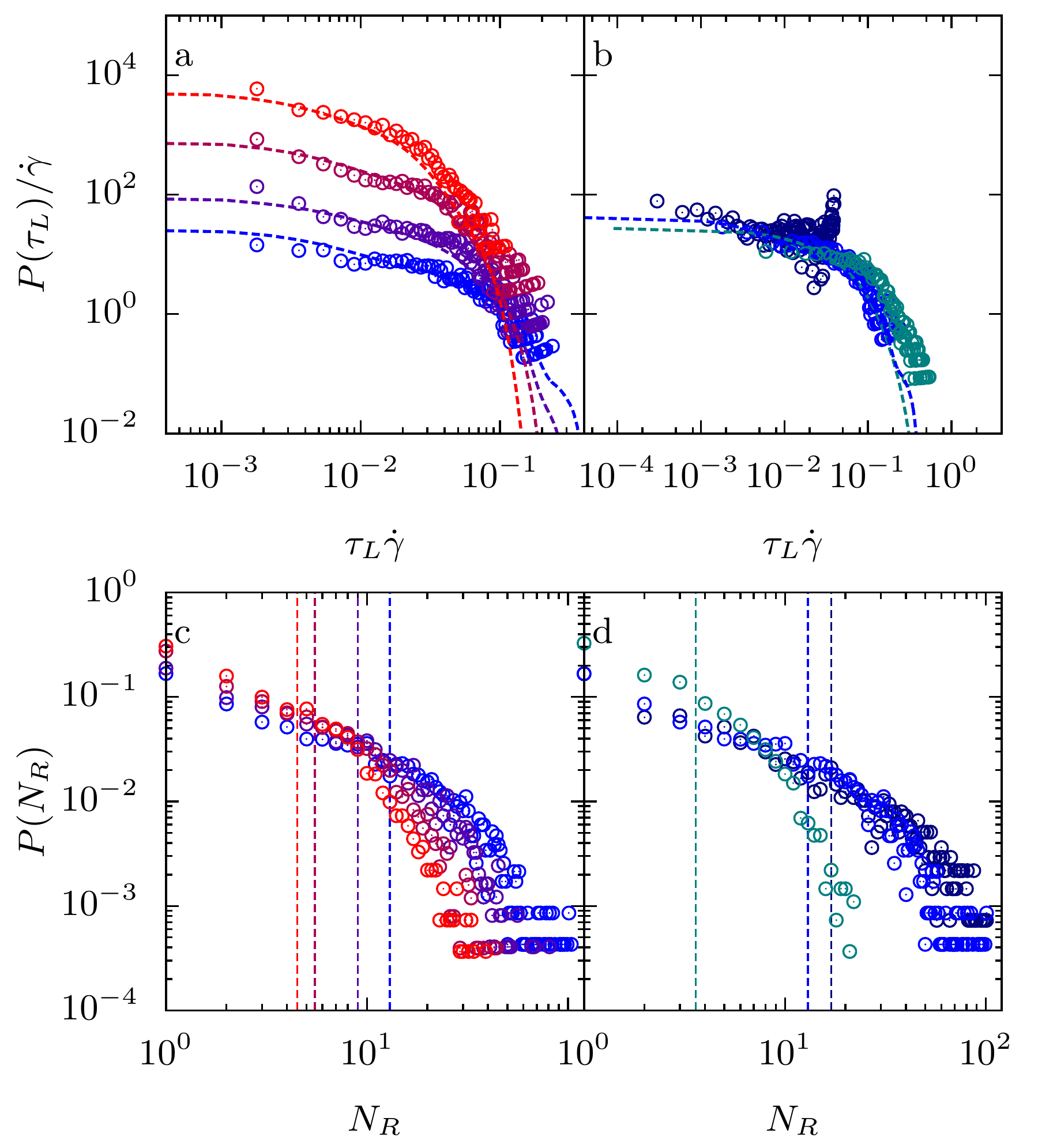}
%\hspace{-0.55pc}\includegraphics[width=0.4925\textwidth]{Rearrangements-2.pdf}\vspace{-0.5pc}
\caption{(a)-(b) Probability distributions for single-soft-spot lifetimes at different temperatures and strain rates respectively. In (a) temperatures of $T=0.1$ (blue) to $T=0.4$ (red) are shown at a strain rate of $\dot\gamma = 10^{-4}$. The lifetime distributions have been shifted vertically for clarity. Overlaid in dashed lines are the predictions of the discrete model. In (b) we show lifetime distributions at strain rates of $\dot\gamma = 10^{-5}$ (dark blue) to $\dot\gamma = 10^{-3}$ (green) at a temperature of $T=0.1$. Again, predictions from the discrete model are shown in dashed lines using for strain rates of $\dot\gamma = 10^{-4}$ and $\dot\gamma = 10^{-3}$. Figures (c)-(d) show the probability distributions for the number of rearrangements needed to destroy a single soft spot using the same color scheme as in (a)-(b). In all cases we see that the number of rearrangements appears to be power-law distributed with an exponential tail. For each distribution the mean number of rearrangements needed to destroy a soft spot is overlaid in dashed line.}
\label{fig:lifetimes}
\centering
\end{figure}

We plot the distribution of soft spot lifetimes, $P(\tau_L)$, in fig.~\ref{fig:lifetimes} (a)-(b) at different temperatures and strain rates respectively. In fig.~\ref{fig:lifetimes} (a) we have shifted the distributions vertically for clarity. In each case we see that lifetimes appear to be power-law distributed up to timescales commensurate with the $\alpha$-relaxation time of the system; at longer time scales, the distribution of lifetimes decays exponentially. Referring to fig.~\ref{fig:lifetimes} (a) we see that the crossover from power-law distributed to exponential distributed lifetimes shifts to shorter times as a function of increasing temperature, as expected for the $\alpha$-relaxation time. This shift is mimicked in the shift of the decay of the $C_{SS}(\delta t)$ function. Considering fig.~\ref{fig:lifetimes} (b) we see that the distribution $P(\tau_L)$ appears to collapse under the mapping $\tau_L\to\tau_L\dot\gamma$ and $P(\tau_L)\to P(\tau_L)/\dot\gamma$ which is again seen in the collapse of the time-dependent correlation functions. Referring to fig.~\ref{fig:correlation} (b) we also see that at the lowest strain rate of $\dot\gamma = 10^{-5}$, our simulation timescale is significantly less than $\tau_\alpha$. This is reflected in distribution of single-soft-spot lifetimes in fig.~\ref{fig:lifetimes} (b) where we see no crossover of the lifetime distribution.

We now argue that the soft spot autocorrelation function follows from the soft spot lifetime distribution introduced above. To do this we assume that when a soft spot is destroyed it is replaced at random somewhere in the system. It follows that the $C_{SS}(\delta t)$ function measures the fraction of soft spots that have not yet decayed after a time $\delta t$. If we additionally assume that soft spots are destroyed at a rate that is independent of their size then this implies that 
\begin{equation}
C_{SS}(\delta t)\sim P(\tau_L\geq \delta t) = 1 - \int_{0}^{\delta t}P(\tau_L)d\tau_L.
\end{equation}
We test this relationship by comparing the measured $C_{SS}(\delta t)$ function (drawn using solid lines) with $P(\tau_L\geq \delta t)$ (drawn using dashed lines) in fig.~\ref{fig:correlation} (e)-(f) for different temperatures and strain rates respectively. A distinct constant of proportionality is used to scale each set data. In fig.~\ref{fig:lifetimes} (d), motivated by the collapsed of $P(\tau_L)$ in fig.~\ref{fig:lifetimes} (b), a single cumulative distribution of soft spot lifetimes was constructed by aggregating soft spot lifetimes from all three strain rates.

The agreement between $C_{SS}(\delta t)$ and $P(\tau_L\geq\delta t)$ is excellent. In each case, $P(\tau_L\geq\delta t)$ decays slightly more slowly than the $C_{SS}(\delta t)$ function. This slower decay is consistent with the fact that $\tau_L$ represents an upper bound on the actual lifetime of soft spots in the configuration. Overall this agreement has several interesting implications for the soft spot picture. First, the behavior of the entire soft spot field can be accurately reduced to the dynamics of individual soft spots. This is the second key result of the paper. Furthermore, the validity of the relationship, $C_{SS}(\delta t)\sim P(\tau_L\geq \delta t)$, means that correlations between successive rearrangements are relatively unimportant for the overall decorrelation of the soft spot field.  In addition, the degree of agreement suggests that our assumption that the lifetime of individual soft spots is independent of soft spot size is reasonably accurate. Were either of these two assumptions strongly violated, the soft spot autocorrelation function would not be so simply related to the cumulative distribution of soft spot lifetimes.

To gain more insight into the dynamics of individual soft spots, we now consider the number of rearrangements that must occur near a soft spot to destroy it. To this end, we calculate an upper bound on the number of rearrangements that overlap with a soft spot using the $C_{S_\alpha S_\alpha}(\delta t)$ function defined in eq.~\ref{eq:single}. In particular, recall that we associate a soft spot $\alpha$, constructed at a time $t$, with a soft spot $\beta$ at a time $t+\delta t$ that maximizes the function $\langle \tilde S_{\alpha,i}(t)\tilde S_{\beta,i}(t+\delta t)\rangle.$ Thus, we say that a soft spot $\alpha$ has experienced a rearrangement at a time $t+\delta t$ if at least one particle in the associated soft spot $\beta$ has $D^2_{\text{min}}\gtrsim 15T$. Here the threshold in $D^2_{\text{min}}$ was chosen from the onset of the plateau in fig.~\ref{fig:overlap}. By summing the total number of rearrangements associated with a soft spot before it is destroyed we arrive at an upper bound for the number of rearrangements to destroy a soft spot, $N_R$.

A plot of the distribution of the number of rearrangements needed to eliminate a soft spot, $P(N_R)$, can be seen in fig.~\ref{fig:lifetimes} (c)-(d) for varying temperatures and strain rates respectively. The most striking feature of $P(N_R)$ is that soft spots appear to be able to survive many rearrangements before being destroyed. This seems to suggest that the slow decorrelation of the soft spot field as a whole might be related to this high resilience of soft spots to structural rearrangements. Furthermore, in all cases - as with the distribution of soft spot lifetimes - we see distributions that feature broad power-laws with exponential tails. In fig.~\ref{fig:lifetimes} (c) we see that the crossover shifts to smaller $N_R$ as temperature increases. This leads to the appealing hypothesis that the decrease in $\alpha$-relaxation time is related to the fact that rearrangements at higher temperatures are more effective at destroying soft spots than are rearrangements at lower temperatures. We additionally find that soft spots can survive, on average, from 4.5 rearrangements at the lowest temperature to 13 rearrangements at the highest temperature. In fig.~\ref{fig:lifetimes} (d) we see that the crossover of $N_R$ likewise shifts between strain rates of $\dot\gamma = 10^{-3}$ and $\dot\gamma = 10^{-4}$, however this trend does not continue for the lowest strain rate. This is consistent with the observation that at the lowest strain rate we do not observe the system for a single $\alpha$-relaxation. Therefore, in this regime - as with the distribution of single-soft-spot lifetimes - we are unable to see the crossover to exponential destruction of soft spots. Nonetheless, it appears that the number of soft spot destroying rearrangements increases as a function of increasing strain rate. This is once again reflected in the average number of rearrangements that a soft spot can survive which increases from 3 at the highest strain rate to 17 at the lowest.

To understand the role of $P(N_R)$ in determining soft spot lifetimes we introduce a simple model. Consider a system of $N_S$  ``soft spots'' that each require $r_i$ rearrangements to be destroyed. The $r_i$ are to be drawn random from the measured distribution of $N_R$ at some temperature and strain rate. The model proceeds in discrete steps. At each step the system experiences a rearrangement that is randomly distributed across the $N_S$ soft spots. If - at a given step - a soft spot has experienced $r_i$ rearrangements, it is destroyed and replaced by a new soft spot with $r_i$ drawn at random from the distribution of $N_R$.  We solve this model analytically in the appendix, and show that it relates the distribution of soft spot lifetimes to the distribution of $N_R$ in absence of any spatial correlations. In order to convert relate the timescale in the model to the timescale in our simulations, we measure the average number of rearrangements, $R$, that occur in every interval of $2\tau$ as a function of temperature and strain rate. Thus, a single step in the model is rescaled to a time of $2\tau/R$ in our simulations.

The distribution of $\tau_L$ predicted from this model is shown in dashed overlay in fig.~\ref{fig:lifetimes} (a)-(b). In all cases we see fairly good agreement between the measured lifetime distributions and the distribution extracted from the model; in particular we see that the model correctly predicts the initial power-law behavior, the location of the crossover and the exponential tail of the distribution. In fig.~\ref{fig:lifetimes} (a) we see that the model correctly predicts the shift in the crossover as a function of temperature. In fig.~\ref{fig:lifetimes} (b) we show the model predictions only for the two faster strain rates as the distribution of $N_R$ is incomplete for the slowest strain rate. In both cases we see that the analytic lifetime distributions approximately collapse as expected. The success of this model suggests that the number of rearrangements needed to destroy a soft spot is the single most important parameter in reconstructing soft spot lifetimes and hence the entire soft spot autocorrelation function. 

It follows that the soft spot lifetimes, and in turn the behavior of the soft spot field as a whole, can be derived from the distribution of the number of rearrangements necessary to destroy a soft spot. This suggests that the shift in the $\alpha$-relaxation time of the system as a function of temperature is related to the shift in the crossover in the distribution of $N_R$. Thus, at higher temperatures it appears that rearrangements that destroy soft spots become more common. Furthermore, it is apparent that spatial correlations between rearrangements, soft spots, and soft spot sizes appear to be unimportant in predicting the dynamics of the soft spot field as a whole. 

\section{Discussion}

We have shown that soft spots correlate with rearrangements in sheared glasses over a range of temperatures and strain rates. By exploring temperatures ranging from those deep in the glassy phase to those well into the super cooled fluid regime we have shown these correlations to be robust; even at the highest temperature considered, we find that rearrangements are about twice as likely to occur at soft spots than they would be if the soft spots were uncorrelated with rearrangements. Moreover, by showing that soft spots continue to describe plasticity in the supercooled liquid regime we provide evidence that soft spots correlate with dynamical heterogeneities in sheared supercooled liquids. It would be very interesting to study the correlation between soft spots and the enduring displacements that have been observed in the supercooled and glassy regimes~\cite{keys11}, and to see if configurations created using the s-ensemble have a lower density of soft spots~\cite{keys13}.

The decorrelation of the soft spot field has been shown to be extremely slow, featuring correlations lasting as long as the longest timescale for structural relaxation in the system. This slow decay in the correlation function is mirrored in the decorrelation of plastic activity in the system. The strong correlation of soft spots with plasticity in glasses along with the exceptionally long lifetimes of these correlations implies that soft spots are deeply ingrained, long lived structural properties of glassy materials that are in many ways analogous to topological defects in crystalline solids. Finally, we have demonstrated that the behavior of the soft spot field as a whole can be successfully understood in terms of a population of individual soft spots.  In particular, we obtained the surprising result that the soft spot field - and hence plasticity in glassy systems - decorrelates so slowly because many rearrangements are generally required to destroy a single soft spot.

We now speculate that soft spots are robust features within a metabasin~\cite{appignanesi06,vogel04}.   Earlier we showed that the distribution of soft spot lifetimes features a power law tail, and argued that this feature implies that a single rearrangement does not suffice to destroy a single soft spot. Since soft spots are constructed from quasi-localized low-frequency vibrational modes, which have low energy barriers to rearrangement~\cite{xu10}, this suggests that most adjacent minima might feature not only similar soft spots, but similar low barriers to rearrangement.  These minima might correspond to inherent structures within the same metabasin.

If the above characterization of intra-metabasin rearrangements is correct, transitions between the largest metabasins would be marked by the most significant changes to the soft spot field of the inherent structure. This picture would suggest that transitions between the largest metabasins correspond to the annihilation and creation of soft spots.  This would imply a relation between the distribution of soft spot lifetimes and the distribution of inter-metabasin barrier heights in the potential energy landscape.  Testing these speculations would be an interesting avenue for further work.

The success of soft spots in describing rearrangements in sheared systems at nonzero temperatures provides strong support for the hypothesis that soft spots are flow defects in amorphous materials, analogous to topological defects such as dislocations in crystals.  They appear to have some (though not necessarily all) of the properties that have been assumed for shear transformation zones~\cite{falk98,falk11} or regions of fluidity~\cite{bocquet09,mansard11}.  In particular, they are localized, they control rearrangements, and their dynamics are correlated with the relaxation time of the system.  Now that we can track individual soft spots as a function of time, it is important to explore their migration statistics and creation and annihilation rates to put such phenomenological theories on a more solid microscopic footing.

\section{Appendix}

We present an analytic solution to the mean field model outlined in section V. We consider a set of $N_S$ soft spots, each described by a number, $r_i$, of rearrangements that a soft spot $i$ may sustain before it is destroyed. The $r_i$ are independently and identically distributed according to some distribution $R$ with probability density function $P_R(N_R)$. The model proceeds in discrete steps, and at each step a soft spot is randomly selected to experience a rearrangement. Once a soft spot has experienced $r_i$ rearrangements then it is destroyed and replaced with a new soft spot with $r_i$ drawn again at random. As each soft spot is identical, we discard the index $i$ and consider, without a loss of generality, soft spot $0$. 

We calculate the probability density function for the distribution of soft spot lifetimes, $P_T(\tau)$. Consider a soft spot in this model that requires $r$ rearrangements to be destroyed. If this soft spot is to have a lifetime $\tau$ then after $\tau$ steps the soft spot must have experienced exactly $r$ rearrangements. At least one of these rearrangements must occur on step $\tau$, but there are no constraints on how the rest of the rearrangements are to be distributed among the remain $\tau-1$ steps. Therefore, the total number of ways that the $r$ rearrangements might be distributed among the $\tau$ steps is ${\tau-1\choose r-1}$ where ${a\choose b}$ are the binomial coefficients. Since the probability of a soft spot rearranging is $1/N_S$ it follows that the probability of a soft spot having a lifetime $\tau$ given that it requires $r$ rearrangements to be destroyed is,
\begin{equation} 
P_T(\tau | r) = {\tau-1\choose r-1}\left(1-\frac1{N_S}\right)^{\tau-r}\left(\frac1{N_S}\right)^r.
\end{equation}
From standard arguments of conditional probability it therefore follows that,
\begin{equation}P_T(\tau) = \sum_{r = 1}^\infty{\tau-1\choose r-1}\left(1-\frac1{N_S}\right)^{\tau-r}\left(\frac1{N_S}\right)^rP_R(r).\end{equation}
Therefore, in the absence of correlations, we are able to relate a distribution of the number of rearrangements required to destroy a soft spot with a distribution of soft spot lifetimes. 

\begin{acknowledgments}
We thank Lisa Manning and Daniel Sussman for instructive discussions.
JR acknowledges
a UBC Killam Faculty Research Fellowship for financial
support. This work was primarily supported by
the UPENN MRSEC, NSF-DMR-1120901.
\end{acknowledgments}

\bibliography{bibliography}
\end{document}